\documentclass[%
 reprint,
 superscriptaddress,
 amsmath,amssymb,
 aps,
 prr
]{revtex4-2}

\usepackage{graphicx}%
\usepackage{dcolumn}%
\usepackage{bm}%
\usepackage{color}
\usepackage{tensor}
\usepackage{mathtools}

\usepackage{derivative,fixdif}
\usepackage{comment}

\usepackage{physics2}
\usephysicsmodule{ab}
\usephysicsmodule{op.legacy}
\usephysicsmodule{braket}
\usephysicsmodule{nabla.legacy}

\usepackage{here}

\usepackage[colorlinks,linkcolor=blue,urlcolor=blue, citecolor=blue]{hyperref}

\usepackage{mymcro}
\usepackage{appendix}

\newcommand{\red}[1]{\textcolor{black}{#1}}

\begin{document}
\title{High-purity single-photon generation based on cavity QED}

\author{Seigo Kikura}
\email{seigokikura00@gmail.com}
\affiliation{Department of Applied Physics, Graduate School of Engineering, The University of Tokyo, 7-3-1 Hongo, Bunkyo-ku, Tokyo 113-8656, Japan}
\affiliation{Faculty of  Science and Engineering, Waseda University, 3-4-1 Okubo, Shinjuku-ku, Tokyo, 169-8585, Japan}
\author{Rui Asaoka}
\email{rui.asaoka@ntt.com}
\affiliation{Computer and Data Science Laboratories, NTT Corporation, Musashino 180-8585, Japan}
\author{Masato Koashi}
\affiliation{Department of Applied Physics, Graduate School of Engineering, The University of Tokyo, 7-3-1 Hongo, Bunkyo-ku, Tokyo 113-8656, Japan}
\affiliation{Photon Science Center, Graduate School of Engineering, The University of Tokyo, 7-3-1 Hongo, Bunkyo-ku, Tokyo 113-8656, Japan}
\author{Yuuki Tokunaga}
\affiliation{Computer and Data Science Laboratories, NTT Corporation, Musashino 180-8585, Japan}

\begin{abstract}
We propose a scheme for generating a high-purity single photon on the basis of cavity quantum electrodynamics (QED).
This scheme employs an atom as a four-level system and the
structure allows the suppression of the re-excitation process due to the atomic decay, which is known to significantly degrade the single-photon purity in state-of-the-art photon sources using a three-level system.
Our analysis shows that the re-excitation probability arbitrarily approaches zero without sacrificing the photon generation probability when increasing the power of a driving laser between two excited states.
This advantage is achievable by using current cavity-QED technologies.
Our scheme can contribute to developing distributed quantum computation or quantum communication with high accuracy.

\end{abstract}

\maketitle

Single-photon sources based on cavity quantum electrodynamics (QED) are of paramount importance for quantum information processing that ranges from quantum computation~\cite{RefWorks:RefID:78-knill2001scheme,PhysRevA.89.022317} and quantum communication~\cite{kimble2008quantum,Covey_2023}.
Notably, photon sources using a $\Lambda$-type three-level system, which have the advantage of controlling the temporal mode of an emitted photon, have been well studied for their performance~\cite{PhysRevA.99.053843,Vasilev_2010,PhysRevA.106.023712,Giannelli_2018,PhysRevA.76.033804,PhysRevResearch.6.013150}.
Several groups have in fact successfully demonstrated such photon sources using ``atomic" systems, e.g., neutral atoms~\red{\cite{doi:10.1126/science.1095232,PhysRevLett.89.067901,hijlkema2007single,Nisbet-Jones_2011,PhysRevLett.123.133602}}, ions~\cite{WOS:000224730800034,Barros_2009, Indistinguishable_photons_from_a_trapped-ion_quantum_network_node,PhysRevA.102.032616,PRXQuantum.2.020331,Ward_2022}, and defects in solids~\cite{sweeney2014cavity,PhysRevLett.129.053603}.

In photonic quantum information science, various protocols harness quantum interference of photons, where photon sources should preferably generate high-purity single photons.
In the three-level photon generation scheme, however, the atomic spontaneous decay and following re-excitation reduce the purity of photons.
In fact, several recent experiments have reported that the re-excitation significantly deteriorates the Hong-Ou-Mandel interference visibility~\cite{Indistinguishable_photons_from_a_trapped-ion_quantum_network_node,PhysRevLett.130.050803,PhysRevA.102.032616}.
Although choosing an appropriate initial state or truncating the generation process has been proposed to tackle this problem~\cite{PhysRevA.102.032616,Barros_2009}, these strategies have limitations in improving the purity, and especially the latter also sacrifices the total photon generation probability.

Here we propose a scheme for generating a high-purity single photon on the basis of cavity QED. 
This scheme employs a four-level system including two excited states, two (meta)stable states, and two driving lasers.
Using an additional energy level to suppress atomic decay has been examined in some protocols~\cite{PhysRevA.78.053816,PhysRevLett.114.110502,PhysRevLett.118.213601,PhysRevA.97.053831}.
In our protocol, this strategy is developed to generate a high-purity photon.
To evaluate the performance of our scheme, we analyze the four-level system proposed here by using the effective operator formalism~\cite{PhysRevA.85.032111}, which shows that the re-excitation probability is greatly inhibited in our scheme compared to in the previous three-level one. 
The distinct advantage of our scheme is that the photon generation probability is hardly sacrificed.
Finally, we propose a realistic implementation of our scheme using current cavity-QED technologies.

\begin{figure}
  \centering
  \includegraphics[width=0.9\linewidth]{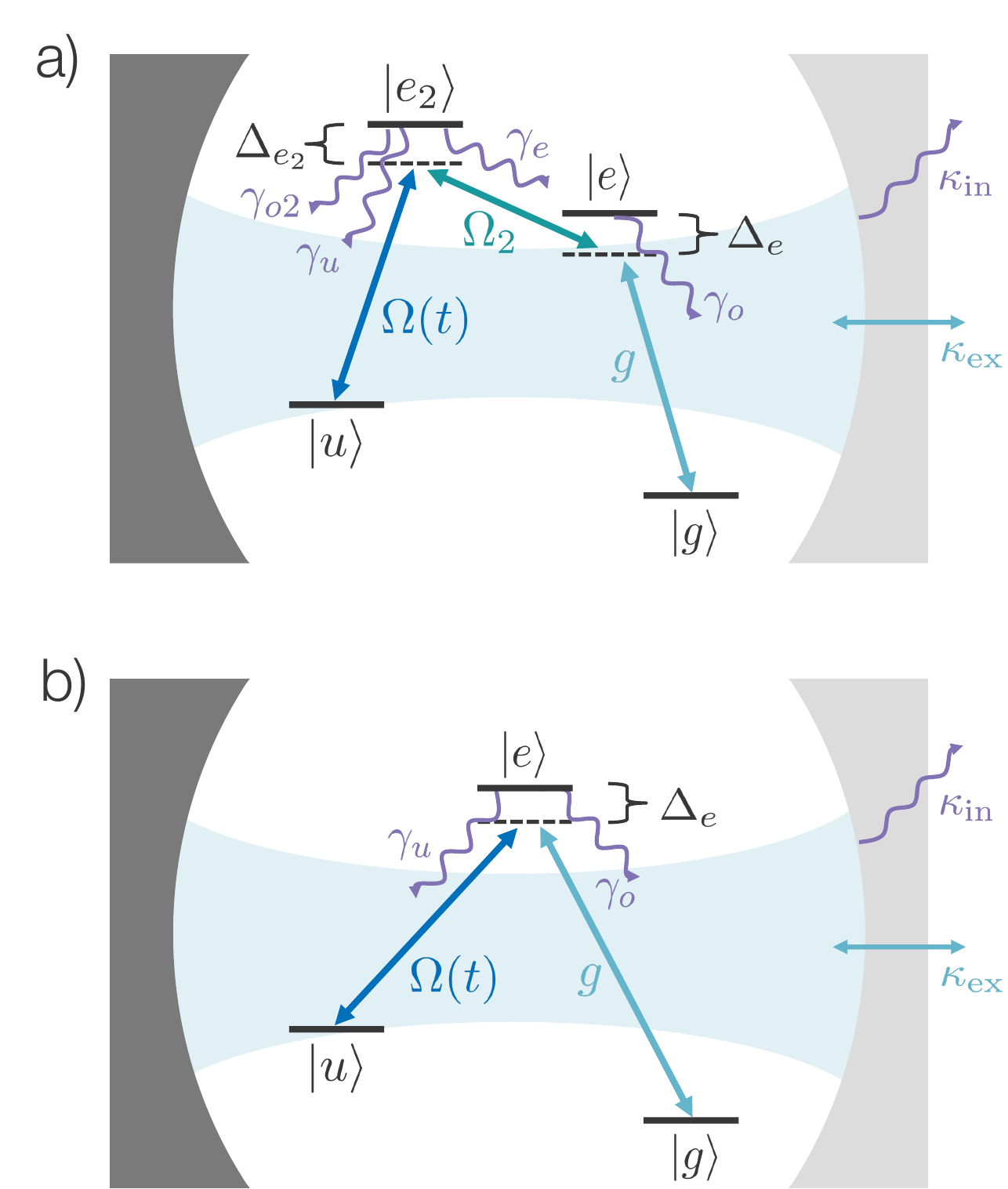}
  \caption{\red{(a) Four-level system inside a cavity. 
  (b) Conventional three-level system for performance comparison with the four-level system.}
  }
  \label{図:4準位図}
\end{figure}
\red{We analyze a four-level atom inside a one-sided cavity shown in Fig.~\ref{図:4準位図}(a) to compare our method with the conventional one that uses a $\Lambda$-type three-level system shown in Fig.~\ref{図:4準位図}(b).
The four-level system is formed by two (meta)stable levels $\ket{u},\ket{g}$ and two excited levels $\ket{e}, \ket{e_2}$.
The $\ket{u}\text{-}\ket{e_2}$ transition is driven by an external driving pulse of frequency $\omega_d$ with Rabi frequency $2\Omega(t)$.
The $\ket{e_2}\text{-}\ket{e}$ transition is driven by a laser of frequency $\omega_2$ with Rabi frequency $2\Omega_2$, which is defined as a real constant.
The $\ket{g}\text{-}\ket{e}$ transition is coupled to a cavity mode of frequency $\omega_c$ with coupling strength $g$, which is defined as a real number.
We prepare the initial state as $\ket{u,0}$, where the Fock state of the cavity mode is denoted as $\ket{0}$ or $\ket{1}$.
In this case, the Hamiltonian in a proper rotating frame is given as 
\begin{equation}
    \begin{split}
        \hat{H} =& \Delta_e \ketbra{e,0}{e,0} + \Delta_{e_2} \ketbra{e_2,0}{e_2,0} \\
        &+ (\Omega(t)\ketbra{e_2,0}{u,0} +\Omega_2\ketbra{e,0}{e_2,0} + \text{h.c.}  ) \\
        &+ (g\ketbra{g,1}{e,0} + \text{h.c.}  ),
    \end{split}
\end{equation}
where $\hat{c}$ is the annihilation operator for the cavity mode, $\omega_l\,(l = u,e,e_2,g)$ is a frequency for the energy of $\ket{l}$, $\Delta_e = \omega_{e} - \omega_u-\omega_d-\omega_2,\Delta_{e_2} = \omega_{e_2}-\omega_u-\omega_d$, and we assume $\omega_u+\omega_d+\omega_2=\omega_g+\omega_c$.
Henceforth, we set $\hbar = 1$.}

\red{
Considering atomic spontaneous decays, we classify them into two groups: decay to levels inside the manifold $\{\ket{u,0}, \ket{e_2,0}, \ket{e,0}, \ket{g,1}\}$ ($\gamma_u,\gamma_{e}$) and levels outside the manifold ($\gamma_{o},\gamma_{o2}$).
The levels outside the manifold may include $\ket{g}$ and other atomic levels except $\ket{u}$ and $\ket{e}$.
Without loss of generality, We treat these as a single (virtual) level $\ket{o}$.
We consider the system where the decay $\ket{e}\to \ket{u}$ is prohibited by a selection rule.}

\red{
Here, we assume $\gamma_{o2}=\gamma_{e}=0$ to compare the four-level system with the three-level one shown in Fig.~\ref{図:4準位図}(b).
We will later discuss the case with $\gamma_{o2}, \gamma_{e} > 0$ and show that these decays will not be an additional problem.
We describe the cavity decay and atomic spontaneous decay with Lindblad operators:
$\hat{L}\td{ex} = \sqrt{2\kappa_{\text{ex}}}\ketbra{g,0}{g,1}$ corresponds to the cavity decay where a photon is emitted into the desired external mode, $\hat{L}\td{in} = \sqrt{2\kappa_{\text{in}}}\ketbra{g,0}{g,1}$ corresponds to scattering into undesired modes and intracavity losses, $\hat{L}_{u} = \sqrt{2\gamma_{u}}\ketbra{u,0}{e_2,0}$ corresponds to the atomic decay from $\ket{e_2}$ to $\ket{u}$, and $\hL_o = \sqrt{2\gamma_o}\ketbra{o,0}{e,0}$ corresponds to the atomic decay from $\ket{e}$ to outside the manifold $\{\ket{u,0}, \ket{e_2,0}, \ket{e,0}, \ket{g,1}\}$.}
We define $2\kappa = 2(\kappa\td{ex}+\kappa\td{in})$ as the total cavity loss rate and $2\gamma = 2(\gamma_u+\gamma_o)$ as the total rate of spontaneous decay.

The master equation describes the time evolution of this atom-cavity system as
\begin{equation}
    \odv{\hrho}{t} = -i[\hH,\hrho] + \sum_{x} \ab(\hat{L}_x \hrho \hat{L}_x\da -\frac{1}{2}\{\hat{L}_x\da \hat{L}_x,\hrho\}),
    \label{量子マスター方程式}
\end{equation}
where $x = \text{ex},\text{in},u,o $.
The quantum jumps $\hat{L}\td{in}$ and $\hat{L}_o$ lead to the failure of the photon generation, whereas the quantum jump $\hat{L}\td{ex}$ leads to the success.
The quantum jump $\hat{L}_u$ initializes the system and then the photon generation process re-starts.
We call a process where the decay $\hat{L}_u$ does not occur a single excitation and a process where the decay $\hat{L}_u$ occurs even just once a re-excitation process.

We assume that the intensity and time variation of the driving pulse are small so that the excited states quickly reach an equilibrium.
This assumption, which has been used in many previous works \cite{doi:10.1080/09500349708231869,PhysRevA.76.033804,Giannelli_2018,PhysRevLett.123.133602}, enables us to use the effective operator formalism \cite{PhysRevA.85.032111}.
This formalism shows that a state in an excited subspace is $\hrho\td{excited} = \ketbra*{\psi\td{excited}}{\psi\td{excited}}$ with $\ket*{\psi\td{excited}} \propto [g^2+\kappa(\gamma_o+i\Delta_e)]\ket{e_2,0}-\Omega_2(i\kappa\ket{e,0}+g\ket{g,1})$, and describes the time evolution in a ground subspace as a master equation with effective Hamiltonian and Lindblad operators 
as follows~\cite{supplemental.material}:
\red{
\begin{align*}
    \hat{H}^{\text{eff}}(t) =& \Delta^{\text{eff}}(t)\ketbra{u,0}{u,0}, \\
    \hat{L}_{\text{ex}}^{\text{eff}}(t) =& k_{\text{ex}}^{}(t) \ketbra{g,0}{u,0},\, \hat{L}_{\text{in}}^{\text{eff}}(t) = k_{\text{in}}(t) \ketbra{g,0}{u,0}, \\
    \hat{L}_{u}^{\text{eff}}(t) =& l_{u}(t) \ketbra{u,0}{u,0},\, \hat{L}_{o}\tu{eff}(t) = l_{o}(t) \ketbra{o,0}{u,0}.
\end{align*}}

Using the effective operators, we determine the final state of the desired mode in the following way~\cite{PhysRevA.102.052614}. 
We classify the single-photon emission events using a label $s$, which is a non-negative number: In the event with $s=0$, no quantum jumps occur during $0<t<t_1$ and a jump $\hat{L}\td{ex}^{\text{eff}}$ occurs at $t=t_1$, where $t_1>0$ is arbitrary.
For $s>0$, the event $s$ refers to the case where a jump $\hat{L}^{\text{eff}}_u$ occurs at $t=s$, no quantum jumps occur during $s<t<t_1$, and a jump $\hat{L}^{\text{eff}}\td{ex}$ occurs at $t=t_1$, where $t_1>s$ is arbitrary.
Those events and the complement (no photon emission) are mutually exclusive and are also
distinguishable without looking at the desired mode in principle because the occurrence and its time of a quantum jump ($\hat{L}^{\text{eff}}\td{in},\hat{L}^{\text{eff}}_u,\hat{L}^{\text{eff}}_o$) can be recorded in the associated environment. 
Hence the final state of the desired mode is given by
a mixture of those events as
\begin{equation}
  \hat{\varrho} = \ketbra{\psi_0}{\psi_0} + \int_0^{\infty}\dd{s} r(s)\ketbra{\psi_s}{\psi_s} + (1-P\td{total})\ketbra{0}{0},
  \label{光子の密度演算子}
\end{equation}
where $\ket{0}$ is a vacuum state of the desired mode and $P\td{total}$ is the total photon generation probability.
Here $r(s)$ is the rate of quantum jump $\hat{L}^{\text{eff}}_u$ at $t=s$, which is to be determined from the effective master equation. 
The unnormalized state vector $\ket{\psi_s}\, (s\ge 0)$ represents the state of the emitted single photon with no quantum jumps during $s<t$ except the jump $\hat{L}^{\text{eff}}\td{ex}$, on condition that the atom-cavity system is in state $\ket{u,0}$ at time $t=s$.
These are pure states because the events involve only quantum jumps $\hat{L}^{\text{eff}}\td{ex}$ whose environment is the desired mode. 
In fact, the state $\ket{\psi_s}$ is related to the time evolution of 
the atom-cavity system as
\begin{equation}
    \ket{g,0}\ket{\psi_s} = \int_s^{\infty} \dd{t} \hat{L}^{\text{eff}}\td{ex}(t) \ha\da(t) \ket{\Phi_s(t)}\ket{0},
\end{equation}
where $\ha(t)$ is the output field operator.
Here $\ket{\Phi_s(t)}$ is the state of the atom-cavity system at time $t$ with no quantum jumps during $s<t$, on condition that it is in state $\ket{u,0}$ at time $t=s$.
More precisely, the normalized solution $\hrho_s(t)$ of the effective master equation with $\hrho_s(s) = \ketbra{u,0}{u,0}$ is written in the form $\hrho_s(t) = \ketbra{\Phi_s(t)}{\Phi_s(t)} + \hrho_{s;\,\text{jumps}}(t)$ with $\hrho_{s;\,\text{jumps}}(t)$ representing an unnormalized state after one or more quantum jumps.
The time dependence of state $\ket{\Phi_s(t)}$ is determined by solving 
\begin{equation*}
    i\odv{\ket{\Phi_s(t)}}{t} = \ab(\hat{H}\tu{eff}(t) - \frac{i}{2}\sum_x \hat{L}_x^{\text{eff}\dagger}(t)  \hat{L}_x\tu{eff}(t)) \ket{\Phi_s(t)}
\end{equation*}
with initial condition $\ket{\Phi_s(s)} = \ket{u,0}$.
The first and second terms on the right-hand side of the Eq.~\eqref{光子の密度演算子} respectively correspond to the single-photon states emitted by single and re-excitation processes.

The probability of the photon generation up to time $t$ for the single(re-) excitation process $P\td{si(re)}(t)$ is calculated as
\begin{gather}
    \begin{aligned}
        P\td{si}(t) =& \int_0^{t} \dd{t\p} |\bra{0}\ha(t\p)\ket{\psi_0}|^2 \\
        =& \frac{\kappa^{\text{eff}}\td{ex}}{\kappa^{\text{eff}} + \gamma^{\text{eff}}} \ab[1- e^{-\int_0^t \dd{t\p} 2(\kappa^{\text{eff}}(t\p) + \gamma^{\text{eff}}(t\p))} ],
    \end{aligned} 
    \label{probability of single excitation in three-level}
    \\
    \begin{aligned}
        P\td{re}(t) =& \int_0^t \dd{t\p} \int_0^\infty \dd{s}  r(s) |\bra{0}\ha(t\p)\ket{\psi_s}|^2 \\
        =& \frac{\kappa^{\text{eff}}\td{ex}}{\kappa^{\text{eff}} + \gamma^{\text{eff}}_o} \ab[1- e^{-\int_0^t \dd{t\p} 2(\kappa^{\text{eff}}(t\p) + \gamma^{\text{eff}}_o(t\p))} ] -P\td{si}(t),
    \end{aligned} \notag
\end{gather}
where 
\begin{gather*}
    \kappa\tu{eff}\td{ex(in)}(t) = \frac{|k\td{ex(in)}(t)|^2}{2} ,\quad \gamma\tu{eff}_{u(o)}(t) = \frac{|l_{u(o)}(t)|^2}{2}, \\
    \kappa\tu{eff}(t) = \kappa\tu{eff}\td{ex}(t) + \kappa\tu{eff}\td{in}(t),\quad \gamma\tu{eff}(t) = \gamma\tu{eff}_{u}(t) + \gamma\tu{eff}_{o}(t).
\end{gather*}
We assume that we apply $\Omega(t)$ for a sufficiently long time so that the population of $\ket{u,0}$ finally becomes zero.
In this case, the photon generation probabilities of the single excitation and total processes are given by
\red{
\begin{align}
    P\td{si} =& \frac{\kappa\td{ex} g^2 \Omega_2^2}{\kappa(g^2+\kappa \gamma_o)\Omega_2^2 + \gamma_u [(g^2+\kappa \gamma_o)^2+\kappa^2 \Delta_e^2]},  \label{4準位系での単一励起過程の光子生成確率} \\
    P\td{total} =& \frac{\kappa\td{ex}}{\kappa}\frac{g^2}{g^2+\kappa\gamma_o},
    \label{4準位系での全体の光子生成確率}
\end{align}}
and thus the ratio of the photon produced by the re-excitation process is given as
\red{
\begin{equation}
    R\td{re} = \frac{\gamma_u[(g^2+\kappa\gamma_o)^2+\kappa^2\Delta_e^2]}{\kappa(g^2+\kappa\gamma_o)\Omega_2^2+\gamma_u[(g^2+\kappa\gamma_o)^2+\kappa^2\Delta_e^2] }.
    \label{4準位系での再励起過程の割合}
\end{equation}}
Note that the total photon generation probability~\eqref{4準位系での全体の光子生成確率} reaches the universal upper bound of photon generation probability in the $\Lambda$-type three-level system \cite{PhysRevA.99.053843}.

The quality of a single-photon source may be evaluated by using several quantities. 
One is the efficiency $P\td{total}$, which has already appeared in the above derivation. 
To further evaluate the quality of the temporal modes of the single photon, we assume that the state of the emitted light is written by
\begin{equation}
    \hat{\varrho} = P\td{total} \hat{\varrho}_S+(1-P\td{total})\ketbra{0}{0}
\end{equation}
and define single-photon purity $D_S$ and single-photon fidelity $F_S$ by
\begin{equation}
    D_S = \Tr[(\hat{\varrho}_S)^2] ,\quad F_S = \braket*[3]{\psi\td{ideal}}{\hat{\varrho}_S}{\psi\td{ideal}},
\end{equation}
where $\ket{\psi\td{ideal}}$ is an ideal single-photon state, which we choose to be $\ket{\psi\td{ideal}} = \ket{\psi_0}/\sqrt{\braket{\psi_0}{\psi_0}}$ here.
This purity $D_S$ relates to the Hong-Ou-Mandel interference visibility~\cite{brańczyk2017hongoumandel}.
Substituting Eq.~\eqref{光子の密度演算子} gives these two measures as~\cite{supplemental.material}
\begin{equation}
    D_S = 1-R\td{re},\quad F_S = 1-\frac{R\td{re}}{2-R\td{re}}.
    \label{単一光子源の性能の評価}
\end{equation}
Therefore, the performance depends only on the total photon generation probability and the ratio of the re-excitation process, rather than the driving pulse.
\red{Note that we also derive the same results in the $\Lambda$-type three-level system~\cite{supplemental.material}.}

The above discussion indicates that the total photon generation probability needs to be improved and the re-excitation process suppressed to improve the performance of the single-photon source.
\red{
In the conventional three-level system, increasing $g$ achieves both goals~\cite{PhysRevA.99.053843}.
This improvement, however, requires an experimental breakthrough.
On the other hand, in our four-level system, increasing $\Omega_2$, which can be easily achieved, reduces the population of $\ket{e_2}$ 
and thus suppresses the re-excitation process because $\ket*{\psi\td{excited}} \propto [g^2+\kappa(\gamma_o+i\Delta_e)]\ket{e_2,0}-\Omega_2(i\kappa\ket{e,0}+g\ket{g,1})$.
More precisely, the re-excitation ratio $R\td{re}$ is minimized for $\Delta_e=0$ and is made to further approach zero by increasing $\Omega_2$ such that $\Omega_2/\sqrt{\gamma_u(g^2/\kappa+\gamma_o)} \gg 1$.
This result also shows that $\Omega_2 \geq g^2/\kappa+\gamma_o$ make the ratio smaller than one of the three-level system $\kappa\gamma_u/(g^2+\kappa\gamma)$~\cite{PhysRevA.99.053843,supplemental.material}.
We stress that the total photon generation probability~\eqref{4準位系での全体の光子生成確率} does not dependent on $\Delta_e$ and $\Omega_2$, and thus our method can reduce the re-excitation process without sacrificing the total photon generation probability.}

\begin{figure}[t]
  \centering
  \includegraphics[width=\linewidth-0.8cm]{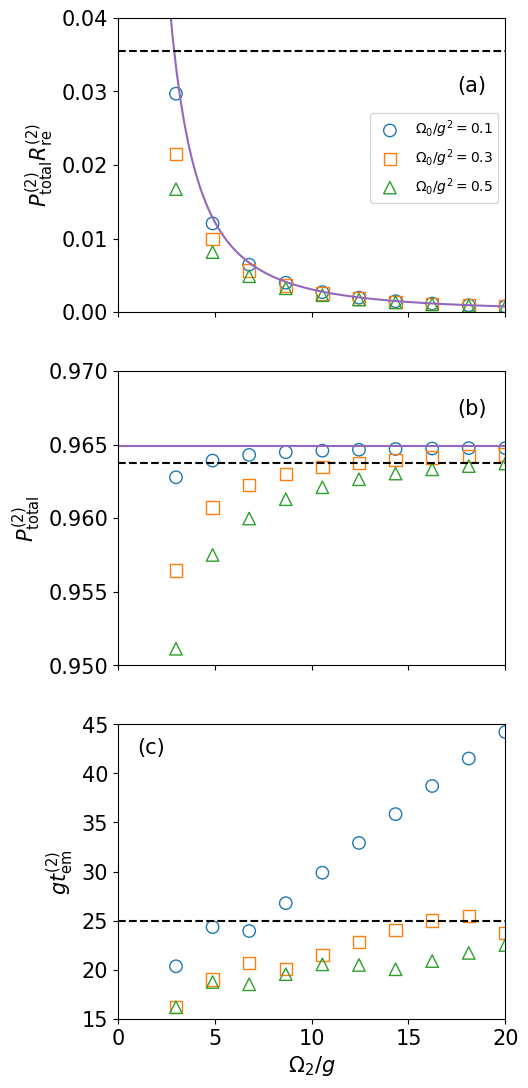}
  \caption{
  (a) Probability of the photon generation via the re-excitation process, (b) total photon generation probability, and (c) photon emission time as a function of $\Omega_2$/$g$.
  All detunings of the three- and four-level systems are zero and $\gamma_u = 0.1g,\gamma_g=\kappa\td{in} = 0.01g$ and  $\kappa\td{ex}= \kappa\td{in}\sqrt{1+g^2/\kappa\td{in}\gamma}$, which maximizes $P\td{si}$~\cite{PhysRevA.99.053843}.
  Dashed lines show the result of the three-level system that includes the minimum emission time when optimizing $\Omega_0$ and the corresponding photon generation probability.
  The optimal value is $\Omega_0/g^2 = 0.070$.
  Solid lines show the analytical result. 
  }
  \label{図:光子生成確率と時間}
\end{figure}
We calculate the photon generation probability and an emission time by numerically solving the master equation \eqref{量子マスター方程式}, where we use a QuTiP 2 package \cite{JOHANSSON20131234}.
In this calculation, we assume the driving pulse that has a linear shape, i.e., $\Omega(t) = \Omega_0 t$.
The symbols in Figs.~\ref{図:光子生成確率と時間}(a) and \ref{図:光子生成確率と時間}(b) show the photon generation probability for the re-excitation process and the total probability with the analytical values represented by the solid lines.
These simulations show that the smaller $\Omega_0$ is, the smaller the deviation between exact solutions and ones by the effective operator formalism is.
These numerical results demonstrate that setting $\Omega_0$ small and increasing $\Omega_2$ such that $\Omega_2/g \geq (g^2/\kappa +\gamma_o)/g =3.2$ suppresses the re-excitation process without diminishing the total photon generation probability compared to the three-level system.

Figure~\ref{図:光子生成確率と時間}(c) shows an emission time by numerical calculation.
Here, we define a photon emission time of the four-level system $t\td{em}$ as a time when the photon generation probability for the single excitation process becomes 0.99 of its maximum value, i.e. $0.99 \times (1-R\td{re})P\td{total}$.
We can find that the photon emission time increases when $\Omega_2$ increases.
This behavior can be explained analytically.
We can find that the generation probability up to time $t$ for the single excitation process is given as $P\td{si}(t) =  (1-R\td{re})P\td{total}(1-e^{-\lambda\td{si}h(0,t)})$, where we define $h(\tau_1,\tau_2) = \int_{\tau_1}^{\tau_2} \dd{t} |\Omega(t)|^2$~\cite{supplemental.material}.
This equation indicates that for a given shape of $\Omega(t)$, increasing $\lambda\td{si}$ makes $t\td{em}$ small.
We can derive the upper bound of $\lambda\td{si}$ as $\lambda\td{si} \leq 2R\td{re}/\gamma_u$ whose equality holds when
\begin{equation}
    \Delta_{e_2} = \frac{\kappa^2\Omega_2^2 \Delta_e}{(g^2+\kappa\gamma_o)^2+\kappa^2\Delta_e^2}.
    \label{4準位，光子生成時間を最小化する条件式}
\end{equation}
This inequality thus shows a trade-off between the photon emission time and the ratio of the re-excitation process.
Note that this trade-off is also valid for the three-level system.
However, the numerical result shows that increasing $\Omega_0$ suppresses the delay of photon generation with a slight sacrifice of $P\td{total}$.

So far, we have assumed $\gamma_{o2}=\gamma_{e}=0$, but there may be non-negligible decay $\ket{e_2} \to \ket{o},\ket{e_2} \to \ket{e}$ in real atoms.
The former decay does not decrease the single-photon purity and fidelity but the total photon generation probability, whereas the latter decay affects all of them.
However, we can neglect both decays in the limit $\Omega_2 \gg \sqrt{(\gamma_u+\gamma_{o2}+\gamma_e)(g^2/\kappa + \gamma_o)}$ because increasing $\Omega_2$ decrease the population of $\ket{e_2}$ and thus suppress all decays from $\ket{e_2}$~\cite{supplemental.material}.

To realize our scheme, we need to use an atom that satisfies the following two requirements: 
(i) The decay $\ket{e}\to \ket{u}$ is prohibited by the selection rule. 
(ii) The transition $\ket{e_2}\text{-}\ket{e}$ can be driven with a \red{sufficient} large $\Omega_2$.
One strategy to realize such a system may be to use one-electron atoms with levels in $S$ or $D$ orbital as $\ket{u}$ and $\ket{g}$ and ones in $P$ orbital as $\ket{e}$ and $\ket{e_2}$~\cite{PhysRevLett.114.110502}.
\red{Driving the transition $\ket{e_2}\text{-}\ket{e}$ with a large Rabi frequency may not be easy because the electric-dipole transition is forbidden.
However, a fast microwave-driven transition has recently been reported~\cite{Weber2024}, and thus the above implementation could be possible.}
\red{An alternative choice is to use two-electron atoms, which have been attracting attention from the viewpoint of utilizing their metastable levels in $P$ orbital for quantum information processing~\cite{PhysRevX.12.021027,PhysRevX.12.021028,PhysRevA.105.052438,PhysRevX.13.041035}. }
For example, $^{171}\text{Yb}$ satisfies both reqirements by defining the states of interest as $\ket{u} = \tensor[^3]{P}{_0}|F=1/2, m_{F}=-1/2\rangle,\ket{e_2} = \tensor[^3]{D}{_1}| {F}=3/2, m_{F}=1/2\rangle,\ket{e} = \tensor[^3]{P}{_1}|{F}=3/2, m_{F}=3/2\rangle$, and $\ket{g} = \tensor[^1]{S}{_0}|{F}=1/2, m_{F}=1/2\rangle$.
All optical transitions between these states and the coupling with a cavity field have been realized experimentally\red{~\cite{Noguchi2011,PhysRevA.86.051404,PhysRevLett.122.143002,Pedrozo_Pe_afiel_2020}, and thus we believe that our protocol could be realized in the near future.}

In conclusion, we have introduced a scheme for generating a high-purity photon on the basis of cavity QED.
This scheme can reduce the probability of the re-excitation process close to zero, without sacrificing the total photon generation probability, by increasing the driving-laser power between the excited states.
Many photonic quantum information protocols have the ability to handle photon loss, e.g., post-selection by photon detection.
Thus, our scheme can significantly increase the performance of such protocols even if the total photon generation probability is low due to imperfections of the cavity-QED system.
Our scheme may increase the difficulty of experiments, but we have proposed a feasible implementation examples with current cavity-QED technologies.
Moreover, one of them can be straightforwardly combined with the scheme of entanglement generation between a $^{171}\text{Yb}$ qubit and a time-bin encoded photon~\cite{Covey2019,PhysRevResearch.3.043154,sunami2024}.
This scheme is based on running a photon generation protocol for each time bin and any atomic decay during those protocols breaks the coherence and ruins the entanglement generation. 
Our proposed photon generation scheme that can suppress atomic decays thus helps to realize high-fidelity entanglement generation, which is a key element for distributed quantum computation and quantum communication.

We thank Shinichi Sunami, Akihisa Goban, and Hiroki Takahashi for their advice.
This research was supported by JST (Moonshot R\&D)(Grant No. JPMJMS2061).

\clearpage
\appendix
\onecolumngrid

\section{Effective operator formalism}
We will summarize the effective operator formalism proposed by Reiter and Sørensen~\cite{PhysRevA.85.032111}.
We consider an open system that consists of two distinct subspaces: ground subspace and decaying excited subspace.
We define $\hat{P}\td{ground}$ and $\hat{P}\td{excited}$ as projection operators onto ground subspace and decaying excited subspace and decompose the Hamiltonian $\hat{H}$ as follows:
\begin{equation}
    \begin{aligned}
        \hH =& \hat{P}\td{ground}\hH \hat{P}\td{ground} + \hat{P}\td{excited}\hH \hat{P}\td{excited} + \hat{P}\td{excited}\hH \hat{P}\td{ground} + \hat{P}\td{ground}\hH \hat{P}\td{excited} \\
        =& \hH\td{ground} + \hH\td{excited} + \hat{V}_+ + \hat{V}_-.
    \end{aligned}
\end{equation}
We assume that $\hH\td{ground}=0$ and the coupling of these two subspaces $\hat{V}_{\pm}$ are perturbative.
We then approximate
\begin{equation}
    \int_0^t \dd{t\p} e^{-i\hat{H}\td{NH} (t-t\p)} \hat{V}_+ \simeq -i\hH\td{NH}^{-1} \hV_+,
\end{equation}
and find that a state in the excited subspace is given by
\begin{equation}
     \hrho\td{excited} =\hat{P}\td{excited}\hrho \hat{P}\td{excited} = \hat{H}\td{NH}^{-1}\hV_+ \hrho\td{ground} \hV_- (\hat{H}\td{NH}^{-1})\da,
\end{equation}
where we define a non-Hermitian Hamiltonian as
\begin{equation}
    \hH\td{NH} = \hH\td{excited} - \frac{i}{2}\sum_k \hL_k\da \hL_k,
    \label{H_NH}
\end{equation}
and $\hrho\td{ground} = \hat{P}\td{ground}\hrho \hat{P}\td{ground}$.
Accordingly, the dynamics of a state in ground subspace is governed by
\begin{equation}
    \begin{aligned}
        \odv{\hrho\td{ground}}{t} =& (i\hV_-\hat{H}\td{NH}^{-1}\hV_+\hrho\td{ground} + \text{h.c.}) + \sum_k \hL_k \hrho\td{excited} \hL_k\da \\
        =& -i[\hH\tu{eff} , \hrho\td{ground}] + \sum_k \ab[\hat{L}_k\tu{eff} \hrho\td{ground} (\hat{L}_k\tu{eff})\da -\frac{1}{2}\ab\{(\hat{L}_k\tu{eff})\da \hat{L}_k\tu{eff} , \hrho\td{ground} \}]
        \label{original effective master equation}
    \end{aligned}
\end{equation}
with effective Hamiltonian and Lindblad operators
\begin{align}
    \hH\tu{eff} =& -\frac{1}{2} \hV_- [\hH\td{NH}^{-1}+ (\hH\td{NH}^{-1})\da ]\hV_+ , \\
    \hat{L}_k\tu{eff} =& \hL_k \hH\td{NH}^{-1} \hV_+ .
\end{align}

We will now extend the original results to consider the case with time-dependent couplings $\hV_{\pm}(t)$.
By assuming that the time variation of $\hV_{+}(t)$ is small in the time scale set by $\hat{H}\td{NH}^{-1}$, we find 
\begin{equation}
    \int_0^t \dd{t\p} e^{-i\hat{H}\td{NH} (t-t\p)} \hat{V}_+(t\p) \simeq -i\hH\td{NH}^{-1} \hV_+(t).
\end{equation}
Accordingly, we straightforwardly extend the derivation in Reiter and Sørensen~\cite{PhysRevA.85.032111}, 
and the state in the excited subspace is given by
\begin{equation}
    \hrho\td{excited} = \hat{P}\td{excited}\hrho \hat{P}\td{excited} = \hH\td{NH}^{-1} \hV_+(t) \hrho\td{ground} \hV_-(t) (\hH\td{NH}^{-1})\da.
    \label{density operator in excited subspace}
\end{equation}
We also obtain the effective master equation as 
\begin{equation}
    \odv{\hrho\td{ground}}{t} = -i[\hH\tu{eff}(t) , \hrho\td{ground}] + \sum_k \ab[\hat{L}_k\tu{eff}(t) \hrho\td{ground} (\hat{L}_k\tu{eff}(t))\da -\frac{1}{2}\ab\{(\hat{L}_k\tu{eff}(t))\da \hat{L}_k\tu{eff}(t) , \hrho\td{ground} \}]
    \label{our effective master equation}
\end{equation}
with effective Hamiltonian and Lindblad operators
\begin{align}
    \hH\tu{eff}(t) =& -\frac{1}{2} \hV_-(t) [\hH\td{NH}^{-1}+ (\hH\td{NH}^{-1})\da ]\hV_+(t) ,  \label{H^eff(t)} \\
    \hat{L}_k\tu{eff}(t) =& \hL_k \hH\td{NH}^{-1} \hV_+(t) \label{L^eff(t)}.
\end{align}
We apply this extended effective master equation to three- and four-level systems in the following.

\subsection{Three-level system}
We analyze the $\Lambda$-type three-level system as shown in Fig.~1(b) in the article.
The Hamiltonian is given as follows:
\begin{equation}
    \hH^{\red{(3)}} = \hH\td{excited}^{\red{(3)}} + \hat{V}_{+}^{\red{(3)}}(t) + \hat{V}_-^{\red{(3)}}(t),
\end{equation}
where
\begin{align}
    \hH\td{excited}^{\red{(3)}} &= \Delta_e \ketbra{e,0}{e,0} + g(\ketbra{e,0}{g,1} + \ketbra{g,1}{e,0}), \\
    \hat{V}_+^{\red{(3)}}(t) &=  \Omega(t)\ketbra{e,0}{u,0} ,\quad \hat{V}_-^{\red{(3)}}(t) = \Omega^*(t)\ketbra{u,0}{e,0},
\end{align}
where we define $\Delta_e = \omega_e-\omega_g-\omega_c$ and assume $\omega_u+\omega_d = \omega_g+\omega_c$.
Substituting into Eqs.~\eqref{H_NH} and~\eqref{density operator in excited subspace} gives 
\begin{equation}
    \begin{aligned}
        \hrho\td{excited}^{\red{(3)}} =& \ketbra*{\psi\td{excited}^{\red{(3)}}}{\psi\td{excited}^{\red{(3)}}}, \\
        \ket*{\psi\td{excited}^{\red{(3)}}} =& \frac{\Omega(t)}{g^2+\kappa(\gamma+i\Delta_e)}  \sqrt{\braket[3]{u,0}{\hrho}{u,0}} (i\kappa \ket{e,0} + g\ket{g,1}).
    \end{aligned}
\end{equation}
From Eqs.~\eqref{H^eff(t)} and \eqref{L^eff(t)}, we can find:
\begin{align}
    \hH^{\red{(3)}\text{eff}}(t) &= - \Delta_e |A^{\red{(3)}}(t)|^2 \ketbra{u,0}{u,0},\\
    \hL^{\red{(3)}\text{eff}}\td{ex(in)}(t) &=  \sqrt{2\kappa\td{ex(in)}}\frac{g}{\kappa} A^{\red{(3)}}(t) \ketbra{g,0}{u,0}, \label{L^eff_ex|3準位}  \\
    \hat{L}_{u}^{\red{(3)}\text{eff}}(t) &= i \sqrt{2\gamma_{u}}A^{\red{(3)}}(t) \ketbra{u,0}{u,0}, \\
    \hat{L}_{o}^{\red{(3)}\text{eff}}(t) &= i \sqrt{2\gamma_{o}}A^{\red{(3)}}(t) \ketbra{o,0}{u,0}, 
\end{align}
where we define
\begin{equation}
    A^{\red{(3)}}(t) = \frac{\kappa}{g^2+\kappa(\gamma+i\Delta_e)}\Omega(t).
\end{equation}
To ease the following notation, we define the effective detuning and decay rates as follows:
\begin{align}
     \Delta^{\red{(3)}\text{eff}}(t) =& - \Delta_e |A^{\red{(3)}}(t)|^2 , \\
    \kappa_{\text{ex(in)}}^{\red{(3)}\text{eff}}(t) =& \frac{1}{2} \ab| \sqrt{2\kappa\td{ex(in)}}\frac{g}{\kappa} A^{\red{(3)}}(t)|^2  = \frac{\kappa\td{ex(in)}g^2 }{\kappa^2} |A^{\red{(3)}}(t)|^2, \\
    \gamma_{u(o)}^{\red{(3)}\text{eff}}(t) =& \frac{1}{2}\ab|i \sqrt{2\gamma_{u(o)}}A^{\red{(3)}}(t)|^2  = \gamma_{u(o)}|A^{\red{(3)}}(t)|^2, \\
    \kappa^{\red{(3)}\text{eff}}(t) =& \kappa_{\text{ex}}^{\red{(3)}\text{eff}}(t) + \kappa_{\text{in}}^{\red{(3)}\text{eff}}(t) ,\quad \gamma^{\red{(3)}\text{eff}}(t) = \gamma_{u}^{\red{(3)}\text{eff}}(t) + \gamma_{o}^{\red{(3)}\text{eff}}(t).
\end{align}

We first derive the state of the atom-cavity system at time $t$ with no quantum jumps during $s<t$, on condition that it is in state $\ket{u,0}$ at time $t=s$.
The time dependence of state $\ket*{\Phi_s^{\red{(3)}}(t)}$ is determined by solving 
\begin{equation}
    i\odv{\ket*{\Phi_s^{\red{(3)}}(t)}}{t} = \ab(\hat{H}^{\red{(3)}\text{eff}}(t) - \frac{i}{2}\sum_x \hat{L}_x^{(3)\text{eff}\dagger}(t)  \hat{L}_x^{\red{(3)}\text{eff}}(t)) \ket*{\Phi_s^{\red{(3)}}(t)}
\end{equation}
with initial condition $\ket*{\Phi_s^{\red{(3)}}(s)} = \ket{u,0}$.
We can find that
\begin{equation}
    \hat{H}^{\red{(3)}\text{eff}}(t) - \frac{i}{2}\sum_x \hat{L}_x^{(3)\text{eff}\dagger}(t)  \hat{L}_x^{\red{(3)}\text{eff}}(t) = [\Delta^{\red{(3)}\text{eff}}(t) -i(\kappa^{\red{(3)}\text{eff}}(t)+\gamma^{\red{(3)}\text{eff}}(t)) ]\ketbra{u,0}{u,0}
\end{equation}
and 
\begin{equation}
    \begin{aligned}
        \ket*{\Phi_s^{\red{(3)}}(t)} =& e^{-\int_s^t \dd{t\p}  (\kappa^{\red{(3)}\text{eff}}(t\p) +  \gamma^{\red{(3)}\text{eff}}(t\p) +i\Delta^{\red{(3)}\text{eff}}(t\p))}\ket{u,0} \\
        =& \exp\ab[-\frac{\kappa}{g^2 + \kappa(\gamma +i\Delta_e)} h(s,t)] \ket{u,0}.
    \end{aligned}
    \label{no-jump system state|three level}
\end{equation}
Equations~\eqref{L^eff_ex|3準位} and~\eqref{no-jump system state|three level} give the emitted single-photon state $\ket*{\psi_s^{\red{(3)}}} = \int_0^{\infty} \dd{t} \psi_s^{\red{(3)}}(t) \ha\da(t)\ket{0}$ as
\begin{align}
    \psi_s^{\red{(3)}}(t) =& 0 \quad \text{if} \quad t < s , \\
    \psi_s^{\red{(3)}}(t) =& \braket[3,big]{g,0}{\hat{L}\td{ex}^{\red{(3)}\text{eff}}(t)}{\Phi_s^{\red{(3)}}(t)} =   \sqrt{2\kappa\td{ex}} \frac{g}{g^2+\kappa(\gamma+i\Delta_e)} \Omega(t) \exp\ab\ab[-\frac{\kappa}{g^2 + \kappa(\gamma +i\Delta_e)} h(s,t)]   \quad  \text{if} \quad t \geq s .
\end{align}
Second, we derive the rate of quantum jump $\hL_u^{\red{(3)}\text{eff}}$ at $t=s$.
From Eq.~\eqref{our effective master equation}, we derive the differential equation for the population of $\ket{u}$ as
\begin{equation}
    \begin{aligned}
        \dot{\rho}_{u,0}^{\red{(3)}} =& -(2\kappa_{\text{ex}}^{\red{(3)}\text{eff}}(t) + 2\kappa_{\text{in}}^{\red{(3)}\text{eff}}(t) + 2\gamma_{o}^{\red{(3)}\text{eff}}(t))\rho_{u,0}^{\red{(3)}} \\
        =& - \frac{2\kappa(g^2+\kappa\gamma_o)}{(g^2+\kappa\gamma)^2+\kappa^2\Delta_e^2} |\Omega(t)|^2 \rho_{u,0}^{\red{(3)}} ,
    \end{aligned}
\end{equation}
where $\rho_{u,0}^{\red{(3)}}  = \braket[3]{u,0}{\hrho^{\red{(3)}}}{u,0}$.
With the initial condition $\rho_{u,0}^{\red{(3)}}(0) = 1$, this equation gives 
\begin{equation}
    \begin{aligned}
        \rho_{u,0}^{\red{(3)}}(t) =& \exp\ab[-\int_0^t \dd{t\p} 2(\kappa^{\red{(3)}\text{eff}}(t\p) + \gamma_{o}^{\red{(3)}\text{eff}}(t\p))  ] \\
        =& \exp\ab[- \frac{2\kappa(g^2+\kappa\gamma_o)}{(g^2+\kappa\gamma)^2+\kappa^2\Delta_e^2} h(0,t) ].
    \end{aligned}
\end{equation}
We then derive the rate $r^{\red{(3)}}(s)$ as follows:
\begin{equation}
    \begin{aligned}
        r^{\red{(3)}}(s) =& \Tr\ab[\hrho^{\red{(3)}}(s) (\hat{L}_{u}^{\red{(3)}\text{eff}}(s))\da \hat{L}_{u}^{\red{(3)}\text{eff}}(s) ] = 2\gamma_{u}^{\red{(3)}\text{eff}}(s) \rho_{u,0}^{\red{(3)}}(s) \\
        =& 2\gamma_{u}^{\red{(3)}\text{eff}}(s) \exp\ab[-\int_0^s \dd{t} 2(\kappa^{\red{(3)}\text{eff}}(t)  + \gamma_{o}^{\red{(3)}\text{eff}}(t))  ] \\
        =& \frac{\kappa^2\gamma_u}{(g^2+\kappa\gamma)^2+\kappa^2\Delta_e^2} |\Omega(s)|^2 \exp\ab[- \frac{2\kappa(g^2+\kappa\gamma_o)}{(g^2+\kappa\gamma)^2+\kappa^2\Delta_e^2} h(0,s) ].
    \end{aligned}
\end{equation}

We can find the probability of the photon generation from those values.
The probability of the photon generation up to time $t$ for the single(re-) excitation process $P\td{si(re)}^{\red{(3)}}(t)$ is given as follows:
\begin{gather}
    \begin{aligned}
        P\td{si}^{\red{(3)}}(t) =& \int_0^t \dd{t\p}  |\psi_0^{\red{(3)}}(t\p)|^2 \\
        =& \int_0^t \dd{t\p} 2\kappa_{\text{ex}}^{\red{(3)}\text{eff}}(t\p) \exp\ab[ -\int_0^{t\p} \dd{t\pp}  2(\kappa^{\red{(3)}\text{eff}}(t\pp) +  \gamma^{\red{(3)}\text{eff}}(t\pp) ) ] \\
        =& \frac{\kappa\td{ex}}{\kappa} \frac{g^2}{g^2+\kappa\gamma} \ab\{ 1 - \exp\ab[- \frac{2\kappa(g^2+\kappa\gamma)}{(g^2+\kappa\gamma)^2+\kappa^2\Delta_e^2} h(0,t)  ]\} \\
        \biggl(=& \frac{\kappa^{\red{(3)}\text{eff}}\td{ex}}{\kappa^{\red{(3)}\text{eff}} + \gamma^{\red{(3)}\text{eff}}} \ab\{1- \exp\ab[-\int_0^t \dd{t\p} 2(\kappa^{\red{(3)}\text{eff}}(t\p) + \gamma^{\red{(3)}\text{eff}}(t\p)) ]  \} \biggr), \label{single-excitation probability|three-level}
    \end{aligned} \\
    \begin{aligned}
         P\td{re}^{\red{(3)}}(t) =&  \int_0^t \dd{t\p} \int_0^\infty \dd{s}  r^{\red{(3)}}(s) |\psi_s^{\red{(3)}}(t\p)|^2 \\
         =& \int_0^t \dd{t\p} |\psi_0^{\red{(3)}}(t\p)|^2 \ab\{\exp\ab(\int_0^{t\p} \dd{t}  2\gamma^{\red{(3)}\text{eff}}\td{u}(t) ) -1 \} \\
         =& \frac{\kappa\td{ex}}{\kappa} \frac{g^2}{g^2+\kappa\gamma_o} \ab\{ 1 - \exp\ab[- \frac{2\kappa(g^2+\kappa\gamma_o)}{(g^2+\kappa\gamma)^2+\kappa^2\Delta_e^2} h(0,t)  ]\} - P\td{si}^{\red{(3)}}(t) \\
         \biggl(=& \frac{\kappa^{\red{(3)}\text{eff}}\td{ex}}{\kappa^{\red{(3)}\text{eff}} + \gamma_o^{\red{(3)}\text{eff}}} \ab\{1- \exp\ab[-\int_0^t \dd{t\p} 2(\kappa^{\red{(3)}\text{eff}}(t\p) + \gamma^{\red{(3)}\text{eff}}_o(t\p) ) ]  \} - P\td{si}^{\red{(3)}}(t)\biggr).
    \end{aligned}
\end{gather}
When applying $\Omega(t)$ for a sufficiently long time so that the population of $\ket{u,0}$ finally becomes zero, the photon generation probabilities of the single excitation and total processes are given as follows:
\begin{align}
    P\td{si}^{\red{(3)}} \coloneq& \lim_{t\to \infty} P\td{si}^{\red{(3)}}(t) = \frac{\kappa\td{ex}}{\kappa}\frac{g^2}{g^2+\kappa\gamma},   \\
    P\td{total}^{\red{(3)}} \coloneq& \lim_{t\to \infty}(P\td{si}^{\red{(3)}}(t)+P\td{re}^{\red{(3)}}(t)) = \frac{\kappa\td{ex}}{\kappa}\frac{g^2}{g^2+\kappa\gamma_o}. 
\end{align}
From Eq.~\eqref{single-excitation probability|three-level}, we can also find 
\begin{equation}
    \lambda\td{si} = \frac{2\kappa(g^2+\kappa\gamma)}{(g^2+\kappa\gamma)^2+\kappa^2\Delta_e^2}.
\end{equation}
This is bounded as
\begin{equation}
    \lambda\td{si} \leq \frac{2R\td{re}}{\gamma_u}
\end{equation}
whose equality holds when $\Delta_e=0$.

\subsection{Four-level system}
We analyze the four-level system as shown in Fig.~1(a) in the article.
We assume $\gamma_{o2}=\gamma_{e}=0$ to compare the four-level system with the three-level one.
We discuss the case with $\gamma_{o2}, \gamma_e > 0$ in Sec.~\ref{app: gamma_o2,gamma_e}.
The Hamiltonian is given by Eq.~(1) in the article and is reproduced here
\begin{align}
    \hat{H}(t) &= \hat{H}\td{excited} + \hat{V}_+(t) + \hat{V}_-(t), \\
    \hH\td{excited} &= \Delta_e \ketbra{e,0}{e,0} + \Delta_{e_2} \ketbra{e_2,0}{e_2,0} + g(\ketbra{e,0}{g,1} + \ketbra{g,1}{e,0}) + \Omega_2 (\ketbra{e,0}{e_2,0}+\ketbra{e_2,0}{e,0} ), \\
    \hat{V}_+(t) &=  \Omega(t)\ketbra{e_2,0}{u,0} ,\quad \hat{V}_-(t) = \Omega^*(t)\ketbra{u,0}{e_2,0}.
\end{align}
By assuming the intensity and time variation of $\Omega(t)$ are small, we can follow the same recipe as before and derive the state in the excited subspace as
\begin{equation}
    \begin{aligned}
        \hrho\td{excited} =& \ketbra*{\psi\td{excited}}{\psi\td{excited}} , \\
        \ket*{\psi\td{excited}} =& \frac{i\Omega(t)}{g^2(\gamma_u+i\Delta_{e_2})+\kappa[(\gamma_o+i\Delta_e)(\gamma_u+i\Delta_{e_2})+\Omega_2^2] }  \sqrt{\braket[3]{u,0}{\hrho}{u,0}}\\
        & \times \{[g^2+\kappa(\gamma_o+i\Delta_e)]\ket{e_2,0}-\Omega_2(i\kappa\ket{e,0}+g\ket{g,1}) \},
    \end{aligned}
    \label{four-level system in excited subspace}
\end{equation}
and the effective operators as follows:
\begin{align}
    \hH^{\text{eff}}(t) &= - \frac{(g^2+\kappa\gamma_o)^2\Delta_{e_2}+\kappa^2(\Delta_e\Delta_{e_2}-\Omega_2^2)\Delta_e }{\kappa^2} |A(t)|^2 \ketbra{u,0}{u,0} , \\
    \hL\td{ex(in)}^{\text{eff}}(t) &= -i \sqrt{2\kappa\td{ex(in)}} \frac{g}{\kappa} \Omega_2 A(t) \ketbra{g,0}{u,0} , \\
    \hL_{u}^{\text{eff}}(t) &= i\sqrt{2\gamma_u} \frac{g^2+\kappa(\gamma_o+i\Delta_e)}{\kappa} A(t) \ketbra{u,0}{u,0} , \\
    \hL_{o}^{\text{eff}}(t) &=  \sqrt{2\gamma_o}\Omega_2 A(t) \ketbra{o,0}{u,0}  ,
\end{align}
where we define
\begin{equation}
    A(t) = \frac{\kappa}{(g^2+\kappa\gamma_o)\gamma_u + \kappa(\Omega_2^2-\Delta_e\Delta_{e_2}) + i[(g^2+\kappa\gamma_o)\Delta_{e_2} + \kappa\gamma_u \Delta_e ] }  \Omega(t).
\end{equation}
We then define the effective detuning and decay rates as follows:
\begin{align}
    \Delta^{\text{eff}}(t) =& - \frac{(g^2+\kappa\gamma_o)^2\Delta_{e_2}+\kappa^2(\Delta_e\Delta_{e_2}-\Omega_2^2)\Delta_e }{\kappa^2} |A(t)|^2, \\
    \kappa_{\text{ex(in)}}^{\text{eff}}(t) =& \frac{1}{2} \ab|-i \sqrt{2\kappa\td{ex}} \frac{g}{\kappa} \Omega_2 A(t)|^2 = \frac{\kappa\td{ex(in)}g^2 \Omega^2 }{\kappa^2} |A(t)|^2, \\
    \gamma_{u}^{\text{eff}}(t) =& \frac{1}{2}\ab|i\sqrt{2\gamma_u} \frac{g^2+\kappa(\gamma_o+i\Delta_e)}{\kappa} A(t) |^2 = \gamma_u \frac{(g^2+\kappa\gamma_o)^2+\kappa^2\Delta_e^2}{\kappa^2} |A(t)|^2  , \\
    \gamma_{o}^{\text{eff}}(t) =& \frac{1}{2}\ab|\sqrt{2\gamma_o}\Omega_2 A(t)|^2 = \gamma_o \Omega^2_2 |A(t)|^2, \\
    \kappa^{\text{eff}}(t) =& \kappa_{\text{ex}}^{\text{eff}}(t) + \kappa_{\text{in}}^{\text{eff}}(t) ,\quad \gamma^{\text{eff}}(t) = \gamma_{u}^{\text{eff}}(t) + \gamma_{o}^{\text{eff}}(t).
\end{align}

From the effective operators, we find the probability of the photon generation up to time $t$ for the single(re-) excitation process $P\td{si(re)}(t)$ as follows:
\begin{align}
    P\td{si}(t) =& \frac{\kappa^{\text{eff}}\td{ex}}{\kappa^{\text{eff}} + \gamma^{\text{eff}}} \ab\{1- \exp\ab[-\int_0^t \dd{t\p} 2(\kappa^{\text{eff}}(t\p) + \gamma^{\text{eff}}(t\p)) ]  \}, \label{single probability|four-level} \\
    P\td{re}(t) =& \frac{\kappa^{\text{eff}}\td{ex}}{\kappa^{\text{eff}} + \gamma_o^{\text{eff}}} \ab\{1- \exp\ab[-\int_0^t \dd{t\p} 2(\kappa^{\text{eff}}(t\p) + \gamma^{\text{eff}}_o(t\p)) ]  \} - P\td{si}(t).
\end{align}
When applying $\Omega(t)$ for a sufficiently long time so that the population of $\ket{u,0}$ finally becomes zero, the photon generation probabilities of the single excitation and total processes are given as follows:
\begin{align}
    P\td{si} =& \frac{\kappa^{\text{eff}}\td{ex}}{\kappa^{\text{eff}} + \gamma^{\text{eff}}} =  \frac{\kappa\td{ex} g^2 \Omega_2^2}{\kappa(g^2+\kappa \gamma_o)\Omega_2^2 + \gamma_u [(g^2+\kappa \gamma_o)^2+\kappa^2 \Delta_e^2]}, \\
    P\td{total} =& \frac{\kappa^{\text{eff}}\td{ex}}{\kappa^{\text{eff}} + \gamma_o^{\text{eff}}} = \frac{\kappa\td{ex}}{\kappa}\frac{g^2}{g^2+\kappa\gamma_o}, 
\end{align}
and the ratio of the re-excitation process is given as
\begin{equation}
    R\td{re} = 1- \frac{P_{\text{si}}}{P_\text{total}} = \frac{\gamma_u[(g^2+\kappa\gamma_o)^2+\kappa^2\Delta_e^2]}{\kappa(g^2+\kappa\gamma_o)\Omega_2^2+\gamma_u[(g^2+\kappa\gamma_o)^2+\kappa^2\Delta_e^2] }.
\end{equation}
From Eq.~\eqref{single probability|four-level}, we can also find
\begin{equation}
    \lambda\td{si} = \frac{2\{\gamma_u[(g^2+\kappa\gamma_o)^2+\kappa^2\Delta_e^2] +(g^2+\kappa\gamma_o)\kappa\Omega_2^2 \}}{[(g^2+\kappa\gamma_o)\gamma_u+\kappa(\Omega_2^2-\Delta_e\Delta_{e_2})]^2+[(g^2+\kappa\gamma_o)\Delta_{e_2}+\kappa\gamma_u\Delta_e]^2}.
\end{equation}
This is bounded as 
\begin{equation}
     \lambda\td{si} \leq \frac{2R\td{re}}{\gamma_u}
\end{equation}
whose equality holds when
\begin{equation}
    \Delta_{e_2} = \frac{\kappa^2\Omega_2^2 \Delta_e}{(g^2+\kappa\gamma_o)^2+\kappa^2\Delta_e^2}.
\end{equation}

\section{Single-photon purity and fidelity}
We use single-photon purity $D_S$ and fidelity $F_S$ as performance measures of a single-photon source.
We first consider the four-level system.
The purity and fidelity are given by
\begin{align}
    D_S =& \frac{1}{P\td{total}^2}\ab(P\td{si}^2 + 2\int_0^\infty \dd{s} r(s)|\braket{\psi_0}{\psi_s} |^2 + \int_0^\infty \dd{s} \int_0^\infty \dd{s\p} r(s)r(s\p) |\braket{\psi_{s\p}}{\psi_s} |^2), \\
    F_S =&  \frac{1}{P\td{total}} \ab(P\td{si} + \frac{1}{P\td{si}}\int_0^{\infty}\dd{s} r(s) |\braket{\psi_0}{\psi_s} |^2) .
\end{align}
If $s\geq s\p$, we can find
\begin{equation}
    \begin{aligned}
     &\braket{\psi_{s\p}}{\psi_s} \\
     =& \exp\ab[\int_0^{s\p} \dd{t\p}  (\kappa\tu{eff}(t\p) +  \gamma\tu{eff}(t\p) -i\Delta\tu{eff}(t\p)  ) ]\exp\ab[\int_0^s \dd{t\p}  (\kappa\tu{eff}(t\p) +  \gamma\tu{eff}(t\p) +i\Delta\tu{eff}(t\p)  ) ]\ab(P\td{si} - P\td{si}(s) ).
    \end{aligned}
\end{equation}
We then derive
\begin{equation}
    \int_0^{\infty}\dd{s} r(s) |\braket{\psi_0}{\psi_s} |^2 = \int_0^{\infty}\dd{s} r(s) \ab(P\td{si} - P\td{si}(s) )^2  \exp\ab[\int_0^s \dd{t\p}  2(\kappa\tu{eff}(t\p) +  \gamma\tu{eff}(t\p)) ]
\end{equation}
and
\begin{equation}
    \begin{aligned}
        &\int_0^\infty \dd{s} \int_0^\infty \dd{s\p} r(s)r(s\p) |\braket{\psi_{s\p}}{\psi_s} |^2 \\
        =& 2 \int_0^\infty \dd{s\p}  r(s\p) \exp\ab[\int_0^{s\p} \dd{t\p}  2(\kappa\tu{eff}(t\p) +  \gamma\tu{eff}(t\p) )]  \int_{s\p}^\infty \dd{s} r(s)\ab(P\td{si} - P\td{si}(s) )^2 \exp\ab[\int_0^{s} \dd{t\p}  2(\kappa\tu{eff}(t\p) +  \gamma\tu{eff}(t\p) )].
    \end{aligned}
\end{equation}
From Eq.~\eqref{single-excitation probability|three-level}, we can derive
\begin{equation}
    \begin{aligned}
        &\int_{s\p}^\infty \dd{s} r(s)\ab(P\td{si} - P\td{si}(s) )^2 \exp\ab[\int_0^{s} \dd{t\p}  2(\kappa\tu{eff}(t\p) +  \gamma\tu{eff}(t\p) )]  \\
        =& P\td{si}^2 \frac{R\td{re}}{2-R\td{re}} \exp\ab[- \int_0^{s\p} \dd{t} 2(2\kappa\tu{eff}(t) +  \gamma\tu{eff}(t) + \gamma\tu{eff}_o(t) )]
    \end{aligned}
\end{equation}
and using this equation gives
\begin{gather}
    \int_0^{\infty}\dd{s} r(s) |\braket{\psi_0}{\psi_s} |^2 = P\td{si}^2 \frac{R\td{re}}{2-R\td{re}}, \\
    \int_0^\infty \dd{s} \int_0^\infty \dd{s\p} r(s)r(s\p) |\braket{\psi_{s\p}}{\psi_s} |^2 = P\td{si}^2  \frac{R\td{re}}{2-R\td{re}} \frac{R\td{re}}{1-R\td{re}}.
\end{gather}
Accordingly, we find
\begin{equation}
    D_S = 1-R\td{re} ,\quad F_S = 1-\frac{R\td{re}}{2-R\td{re}}.
\end{equation}
When considering the three-level system, we can also derive these measures as before:
\begin{equation}
   D_S = 1-R\td{re}^{\red{(3)}} ,\quad F_S = 1-\frac{R\td{re}^{\red{(3)}}}{2-R\td{re}^{\red{(3)}}}.
\end{equation}
These results show that these measures depend only on the ratio of the re-excitation process, rather than $\Omega(t)$.

\section{Validity of our scheme for a wide range of cavity parameters}
\begin{figure}[t]
    \centering
    \includegraphics[width = 12cm]{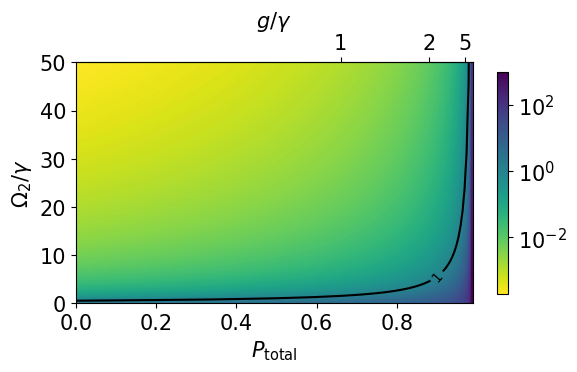}
    \caption{Proportion of the ratio of the re-excitation process for the four-level system to one for the three-level system $R\td{re}/R\td{re}^{\red{(3)}}$. We set $\Delta_e = 0, \gamma_u = \gamma_o, \kappa\td{ex}/\gamma=0.99,\kappa\td{in}/\gamma = 0.01$.}
    \label{fig:ratio_Omega2_and_P_total}
\end{figure}
We show the parameter regime where the four-level scheme is superior to the three-level one. 
Figure~\ref{fig:ratio_Omega2_and_P_total} plots $R\td{re}/R\td{re}^{\red{(3)}}$ as the function of system parameters and the threshold curve $R\td{re}/R\td{re}^{\red{(3)}}=1$. 
When $R\td{re}/R\td{re}^{\red{(3)}} < 1$, the purity of an output photon with the four-level scheme is higher than that with the three-level scheme. 
The values of $g, \gamma_u, \gamma_o, \kappa\td{in}$, and $\kappa\td{ex}$ are shared in both the three-level and four-level cases.
We scan $g/\gamma$, which determines the success probability in both cases ($P\td{total} = P\td{total}^{\red{(3)}}$), and show these values as the horizontal axes.
This result indicates that our scheme is valid for a wide range of $g/\gamma$. 
When $g/\gamma$ becomes enough large, the advantage of our scheme becomes small (in other words, the requirement of $\Omega_2$ becomes high) because the original performance of the cavity is already quite high.

\section{Considering the case with $\gamma_{o2}, \gamma_e \neq 0$}
\label{app: gamma_o2,gamma_e}
So far, we have assumed $\gamma_{o2} = \gamma_e = 0$. We now remove this assumption and analyze these effects.
Our following analytical analysis shows that we can neglect the effect of $\gamma_{o2}$ and $\gamma_{e}$ in the limit $\Omega_2 \gg \sqrt{(\gamma_u+\gamma_{o2} + \gamma_e)(g^2/\kappa + \gamma_o)}$.

We first consider the case with $\gamma_{o2}>0$ and $\gamma_e=0$.
We can describe the decay $\ket{e_2}\to \ket{o}$ by a Lindblad operator $\hat{L}_{o2} = \sqrt{2\gamma_{o2}}\ketbra{o,0}{e_2,0} $.
By using the effective operator formalism, we can find the corresponding effective Lindblad operator as
\begin{equation}
    \hat{L}_{o2}^{\text{eff}} = i\sqrt{2\gamma_{o2}} \frac{g^2+\kappa(\gamma_o+i\Delta_e)}{\kappa} A(t) \ketbra{o,0}{u,0}.
\end{equation}
By replacing $\gamma_o \to \gamma_o + \gamma_{o2}[(g^2+\kappa\gamma_o)^2+\kappa^2\Delta_e^2]/(\kappa^2\Omega_2^2)$, we derive the total photon probability and the ratio of the re-excitation process as
\begin{align}
    P\td{total} &= \frac{\kappa\td{ex}}{\kappa}\frac{g^2}{g^2+\kappa\gamma_o\p}, \label{total probability with gamma_o2} \\
    R\td{re} &=\frac{\gamma_u[(g^2+\kappa\gamma_o\p)^2+\kappa^2\Delta_e^2]}{\kappa(g^2+\kappa\gamma_o\p)\Omega_2^2+\gamma_u[(g^2+\kappa\gamma_o\p)^2+\kappa^2\Delta_e^2] },
\end{align}
where we define
\begin{equation}
    \gamma_o\p = \gamma_o + \frac{(g^2+\kappa\gamma_o)^2+\kappa^2\Delta_e^2}{\kappa^2\Omega_2^2} \gamma_{o2}.
\end{equation}
We can find $ R\td{re}$ is minimized for $\Delta_e = 0$, and in this case $\gamma_o\p \to \gamma_o$ in the limit $\Omega_2 \gg g^2/\kappa + \gamma_o$.

We further consider the case with $\gamma_{o2},\gamma_e>0$.
We can describe the decay $\ket{e_2}\to \ket{e}$ by a Lindblad operator $\hat{L}_{e} = \sqrt{2\gamma_e}\ketbra{e,0}{e_2,0} $.
We now decompose the total atom-cavity density operator as $\hrho(t) = \hrho\td{pure}(t) + \hrho\td{jumps}(t)$ where we define a trajectory with no quantum jumps as $\hrho\td{pure}(t)$.
Note that $\hrho\td{pure}(t)$ does not depend on destinations of quantum jumps.
That is to say, $\hrho\td{pure}(t)$ would be the same if we chose $\hat{L}_{o2} = \sqrt{2\gamma_{o2}}\ketbra{u,0}{e_2,0} $ and $\hat{L}_{e} = \sqrt{2\gamma_e}\ketbra{u,0}{e_2,0} $ instead.
We thus find $\hrho\td{pure,excited} = \hat{P}\td{excited}\hrho\td{pure}\hat{P}\td{excited}$ by replacing $\gamma_u \to \Gamma = \gamma_u+\gamma_{o2}+\gamma_e$ in Eq.~\eqref{four-level system in excited subspace} as
\begin{equation}
    \begin{aligned}
        \hrho\td{pure,excited} =& \ketbra{\phi\td{excited}}{\phi\td{excited}}, \\
        \ket{\phi\td{excited}} =& \frac{i\Omega(t)}{g^2(\Gamma+i\Delta_{e_2})+\kappa[(\gamma_o+i\Delta_e)(\Gamma+i\Delta_{e_2})+\Omega_2^2] }  \sqrt{\braket[3]{u,0}{\hrho\td{pure}}{u,0}}\\
        & \times \{[g^2+\kappa(\gamma_o+i\Delta_e)]\ket{e_2,0}-\Omega_2(i\kappa\ket{e,0}+g\ket{g,1}) \}.
    \end{aligned}
\end{equation}
Using this result gives
\begin{equation}
    \begin{aligned}
        \odv*{\Tr[\hrho\td{pure}(t)]}{t} =& -2\Gamma \rho_{\text{pure};\,e_2,0} -2\gamma_o \rho_{\text{pure};\,e,0} -2\kappa \rho_{\text{pure};\,g,1} \\
        =& -2\frac{\kappa(g^2+\kappa\gamma_o)\Omega_2^2 + \Gamma[(g^2+\kappa\gamma_o)^2+\Delta_e^2]}{g^2\Omega_2^2}\rho_{\text{pure};\,g,1},
    \end{aligned}
    \label{differential equation of trace sigma}
\end{equation}
where we define $\rho_{\text{pure};\,x} = \braket[3]{x}{\hrho\td{pure}}{x}$.
We also find 
\begin{equation}
    \odv{P\td{pure}(t)}{t} = 2\kappa\td{ex} \rho_{\text{pure};\,g,1} ,
    \label{differential equation of Ppure}
\end{equation}
where we define the photon generation probability with no atomic decay up to $t$ as $P\td{pure}(t)$.
Using Eqs~\eqref{differential equation of trace sigma},~\eqref{differential equation of Ppure}, and $\lim_{t\to\infty}\Tr[\hrho\td{pure}(t)] = 0$ gives the photon generation probability with no atomic decay as
\begin{equation}
    P\td{pure} = \frac{\kappa\td{ex} g^2 \Omega_2^2}{\kappa(g^2+\kappa \gamma_o)\Omega_2^2 + \Gamma [(g^2+\kappa \gamma_o)^2+\kappa^2 \Delta_e^2]}.
\end{equation}

We now describe the density operator of the emitted light as
\begin{equation}
    \hat{\varrho} = \hat{\varrho}_0 + \int_0^{\infty} \dd{\tau} 2\gamma_u \rho_{\text{pure};\,e_2,0}(\tau) \hat{\varrho}_1(\tau) + \int_0^{\infty} \dd{\tau} 2\gamma_e \rho_{\text{pure};\,e_2,0}(\tau) \hat{\varrho}_2(\tau) + (1-\bar{P}\td{total})\ketbra{0}{0}.
\end{equation}
Here the first term represents the state of the emitted single photon with no atomic decay.
The second term represents the state of the emitted single photon where no quantum jumps occur until $\tau$ and a jump $\hat{L}_u$ occurs at $\tau$, where $\tau>0$ is arbitrary.
The third term represents the state of the emitted single photon where no quantum jumps occur until $\tau$ and a jump $\hat{L}_e$ occurs at $\tau$, where $\tau>0$ is arbitrary.
We can find $\Tr[\hat{\varrho}_i(\tau)] \leq \kappa\td{ex}/\kappa \,(i=1,2)$, and thus derive 
\begin{equation}
    \begin{aligned}
        &\Tr\ab[\int_0^{\infty} \dd{\tau} 2\gamma_u \rho_{\text{pure};\,e_2,0}(\tau) \hat{\varrho}_1(\tau) + \int_0^{\infty} \dd{\tau} 2\gamma_e \rho_{\text{pure};\,e_2,0}(\tau) \hat{\varrho}_2(\tau)] \\
        \leq& \frac{\kappa\td{ex}}{\kappa} \int_0^{\infty} \dd{\tau} 2(\gamma_u+\gamma_e) \rho_{\text{pure};\,e_2,0}(\tau) \\
        =& \frac{\kappa\td{ex}}{\kappa} \frac{\int_0^{\infty} \dd{\tau} 2(\gamma_u+\gamma_e) \rho_{\text{pure};\,e_2,0}(\tau)}{\int_0^{\infty} \dd{\tau} (2\Gamma \rho_{\text{pure};\,e_2,0}(\tau) +2\gamma_o \rho_{\text{pure};\,e,0}(\tau) +2\kappa \rho_{\text{pure};\,g,1}(\tau))} \\
        =& \frac{\kappa\td{ex}}{\kappa} \frac{(\gamma_u+\gamma_e)[(g^2+\kappa \gamma_o)^2+\kappa^2 \Delta_e^2]}{\kappa(g^2+\kappa \gamma_o)\Omega_2^2 +\Gamma [(g^2+\kappa \gamma_o)^2+\kappa^2 \Delta_e^2]},
    \end{aligned}
\end{equation}
where we use $\int_0^{\infty} \dd{\tau} (2\Gamma \rho_{\text{pure};\,e_2,0}(\tau) +2\gamma_o \rho_{\text{pure};\,e,0}(\tau) +2\kappa \rho_{\text{pure};\,g,1}(\tau))=1$.
This upper bound is minimized for $\Delta_e = 0$ and is made to further approach zero by increasing $\Omega_2$ such that $\Omega_2/\sqrt{\Gamma(g^2/\kappa + \gamma_o)} \gg 1$.

Therefore, we can only produce the photon with no atomic decay and neglect the effect of $\gamma_{o2}$ and $\gamma_{e}$ in the limit $\Omega_2 \gg \sqrt{\Gamma(g^2/\kappa + \gamma_o)}$.

\bibliography{myrefs}

\end{document}